\documentclass[preprint,aps,12pt,showpacs,nofootinbib,tightenlines]{revtex4}
\usepackage{amsmath}
\usepackage{amssymb}
\usepackage{epsfig}
\usepackage{graphicx}
\begin{document}
\preprint{ZJOU-PHY-TH-07-01}
\preprint{NJNU-TH-07-05}

%%%%%%%%%%%%%%%%%%%%%%%%%%%%%%%%%%%%%%%%%%%%%

\newcommand{\beq}{\begin{eqnarray}}
\newcommand{\eeq}{\end{eqnarray}}
\newcommand{\non}{\nonumber\\ }

\newcommand{\acp}{{\cal A}_{CP}}
\newcommand{\etap}{\eta^{(\prime)} }
\newcommand{\etapr}{\eta^\prime }
\newcommand{\pb}{\phi_{B_s}}
\newcommand{\pp}{\phi_{\pi}}
\newcommand{\pe}{\phi_{\eta}^A}
\newcommand{\pepr}{\phi_{\eta'}^A}
\newcommand{\ppp}{\phi_{\pi}^P}
\newcommand{\pep}{\phi_{\eta}^P}
\newcommand{\peprp}{\phi_{\eta'}^P}
\newcommand{\ppt}{\phi_{\pi}^t}
\newcommand{\pet}{\phi_{\eta}^T}
\newcommand{\peprt}{\phi_{\eta'}^T}
\newcommand{\fb}{f_{B_s} }
\newcommand{\fpi}{f_{\pi} }
\newcommand{\feta}{f_{\eta} }
\newcommand{\fetap}{f_{\eta'} }
\newcommand{\rpi}{r_{\pi} }
\newcommand{\re}{r_{\eta} }
\newcommand{\rep}{r_{\eta'} }
\newcommand{\mb}{m_{B_s} }
\newcommand{\mop}{m_{0\pi} }
\newcommand{\moe}{m_{0\eta} }
\newcommand{\moep}{m_{0\eta'} }

\newcommand{\psl}{ p \hspace{-1.8truemm}/ }
\newcommand{\nsl}{ n \hspace{-2.2truemm}/ }
\newcommand{\vsl}{ v \hspace{-2.2truemm}/ }
\newcommand{\epsl}{\epsilon \hspace{-1.8truemm}/\,  }

\def \epjc{ Eur.Phys.J. C }
\def \jpg{  J. Phys. G }
\def \npb{  Nucl. Phys. B }
\def \plb{  Phys. Lett. B }
\def \pr{  Phys. Rep. }
\def \prd{  Phys. Rev. D }
\def \prl{  Phys. Rev. Lett.  }
\def \zpc{  Z. Phys. C  }
\def \jhep{ J. High Energy Phys.  }

%%%%%%%%%%%%%%%%%%%%%%%%%%%%%%%%%%%%%%%%%%%%%%%%%%%%
%%
\title{Studies of $B_s^0 \to \eta^{(\prime)} \eta^{(\prime)}$ decays in the pQCD approach}
\author{Xin Liu$^{a}$\footnote{ liuxin@zjou.edu.cn},
Zhen-Jun Xiao$^{b}$\footnote{ xiaozhenjun@njnu.edu.cn}, Hui-Sheng Wang$^{c}$ }
\affiliation{ $a.$ Department of Physics, Zhejiang
Ocean University, Zhoushan, Zhejiang 316000, P.R. China }
\affiliation{ $b.$ Department of Physics and Institute of Theoretical
Physics, Nanjing Normal University, Nanjing, Jiangsu 210097, P.R. China}
\affiliation{ $c.$ Department of Applied Mathematics and
Physics, Anhui University of Technology and Science, Wuhu, Anhui 241000, P.R. China }  %%
\date{\today}
\begin{abstract}
We calculate the CP averaged branching ratios and CP-violating
asymmetries for $B_s^0 \to \eta \eta, \eta \eta^\prime$ and
$\eta^\prime \eta^\prime$ decays in the perturbative QCD (pQCD)
approach here. The pQCD predictions for the CP-averaged branching ratios are
$Br(B_s^0 \to \eta \eta) = \left (14.2^{+18.0}_{-7.5} \right ) \times 10^{-6}$,
$Br(B_s^0 \to \eta \eta^\prime)= \left ( 12.4 ^{+18.2}_{-7.0} \right ) \times
10^{-6}$, and $Br(B_s^0 \to \eta^{\prime} \eta^{\prime}) =
\left ( 9.2^{+15.3}_{-4.9} \right ) \times 10^{-6}$,
which agree well with those obtained by employing the QCD factorization
approach and also be consistent with available experimental upper limits.
The gluonic contributions are small in size: less than $7\%$
for $B_s \to \eta \eta$ and $ \eta \eta^\prime$ decays, and around $18\%$
for $B_s \to \eta' \eta'$ decay.
The  CP-violating asymmetries for three decays are very small: less than $3\%$ in magnitude.
\end{abstract}

\pacs{13.25.Hw, 12.38.Bx, 14.40.Nd}

\maketitle

Among various  $B \to M_1 M_2$ decay channels ( here $M_i$ refers to the light
pseudo-scalar or vector mesons ), the decays
involving the isosinglet $\eta$ or $\etapr$ mesons in the final
state  are  phenomenologically very interesting and have been studied extensively during the
past decade because of the so-called $K \etapr $ puzzle or other special features
\cite{bn03b,sunbs03,ekou,li0609}.

Motivated by the large number of $B_s$ production and decay events expected at the
forthcoming LHC experiments, the studies about the $B_s$ meson decays become
more attractive than ever before.
Very recently, some two-body $B_s \to M_i \etap$ decays, such as $B_s \to (\pi, \rho, \omega, \phi )
\etap $ decays have been studied in Refs.~\cite{xiao061,xiao062} in the perturbative QCD (pQCD )
factorization approach \cite{cl97,li2003,lb80}.
In this paper, we would like to calculate the branching ratios and CP asymmetries for the
three $B_s^0 \to \eta \eta, \eta \etapr$ and $\etapr \etapr$
decays by employing the low energy effective Hamiltonian
\cite{buras96} and the pQCD approach.
Besides the usual factorizable contributions, we here are able to
evaluate the non-factorizable and the annihilation contributions to these decays.

On the experimental side, only the poor upper limit on $Br(B_s^0
\to \eta \eta)$ is available now \cite{pdg06} (upper limits at
$90\%$ C.L.): \beq Br(B_s^0 \to \eta \eta)< 1.5 \times 10^{-3} \ \
, \label{eq:exp1} \eeq Of course, this situation will be improved
rapidly when LHC experiment starts to run at the end of 2007.

This paper is organized as follows. In Sec.~\ref{sec:f-work}, we
calculate analytically the related Feynman diagrams and present the various
decay amplitudes for the studied decay modes. In Sec.~\ref{sec:n-d}, we show the numerical results
for the branching ratios and CP asymmetries of $B_s^0 \to \eta^{(\prime)} \eta^{(\prime)}$
decays. A short summary and some discussions are also included in this section.

\section{Perturbative calculations}\label{sec:f-work}

Since the b quark is rather heavy we consider the $B_s$ meson at
rest for simplicity. It is convenient to use light-cone coordinate
$(p^+, p^-, {\bf p}_T)$ to describe the meson's momenta: $p^\pm
=(p^0 \pm p^3)/\sqrt{2}$ and ${\bf p}_T = (p^1, p^2)$.
Using the light-cone coordinates the $B_s$ meson and the two final
state meson momenta can be written as
\beq
P_1 =\frac{M_{B_s}}{\sqrt{2}} (1,1,{\bf 0}_T), \quad
P_2 =\frac{M_{B_s}}{\sqrt{2}} (1,0,{\bf 0}_T), \quad
P_3 =\frac{M_{B_s}}{\sqrt{2}} (0,1,{\bf 0}_T),
\eeq
respectively, here the light meson masses have been
neglected. Putting the light (anti-) quark momenta in $B_s$,
$\eta^\prime$ and $\eta$ mesons as $k_1$, $k_2$, and $k_3$,
respectively, we can choose
\beq
k_1 = (x_1 P_1^+,0,{\bf k}_{1T}), \quad k_2 = (x_2 P_2^+,0,{\bf k}_{2T}), \quad k_3 = (0, x_3
P_3^-,{\bf k}_{3T}).
\eeq
Then, after the integration over $k_1^-$, $k_2^-$, and $k_3^+$,  the
decay amplitude for $B_s \to \eta \eta^\prime$ decay, for example,
can be conceptually written as
\beq
{\cal A}(B_s \to \eta \eta^\prime) &\sim &\int\!\! d x_1 d x_2 d x_3 b_1 d b_1 b_2 d b_2
b_3 d b_3 \non && \cdot \mathrm{Tr} \left [ C(t)
\Phi_{B_s}(x_1,b_1) \Phi_{\eta^\prime}(x_2,b_2) \Phi_{\eta}(x_3,
b_3) H(x_i, b_i, t) S_t(x_i)\, e^{-S(t)} \right ], \label{eq:a2}
\eeq
where $k_i$ are momenta of light quarks included in each meson, term $\mathrm{Tr}$ denotes
the trace over Dirac and color indices,
$C(t)$ is the Wilson coefficient evaluated at scale $t$,
the function $H(k_1,k_2,k_3,t)$ is the hard part and can be  calculated
perturbatively, the function $\Phi_M$ is the wave function,
the function $S_t(x_i)$ describes the threshold resummation
~\cite{li02} which smears the end-point singularities on $x_i$, and the
last term, $e^{-S(t)}$, is the Sudakov form factor which suppresses the soft dynamics
effectively. We will calculate analytically the function
$H(x_i,b_i,t)$ for the considered decays in the first order in
$\alpha_s$ expansion and give the convoluted amplitudes in next
section.

For the two-body charmless $B_s$ meson decays, the related weak
effective Hamiltonian $H_{eff}$ can be written as \cite{buras96}
\beq
\label{eq:heff} {\cal H}_{eff} = \frac{G_{F}} {\sqrt{2}} \,
\left[ V_{ub} V_{uq}^* \left (C_1(\mu) O_1^u(\mu) + C_2(\mu)
O_2^u(\mu) \right) - V_{tb} V_{tq}^* \, \sum_{i=3}^{10} C_{i}(\mu)
\,O_i(\mu) \right] \; ,
\eeq
where $C_i(\mu)$ are Wilson coefficients at the renormalization scale $\mu$ and $O_i$ are the
four-fermion operators for the case of $b \to q $ ($q=d,s$) transition \cite{buras96,xiao061}.
For the Wilson coefficients $C_i(\mu)$ ($i=1,\ldots,10$), we will use the leading order (LO)
expressions, although the next-to-leading order (NLO)  results
already exist in the literature ~\cite{buras96}. This is the
consistent way to cancel the explicit $\mu$ dependence in the
theoretical formulae. For the renormalization group evolution of
the Wilson coefficients from higher scale to lower scale, we use
the formulae as given in Ref.\cite{luy01} directly.

\subsection{Decay amplitudes}

We firstly  take  $B_s \to \eta \eta' $ decay mode as an example, and
then extend our study to $B_s \to \eta \eta$ and $ \eta^\prime \eta^\prime$ decays.
Similar to the $B_s^0 \to \pi^0 \etap$ decays in \cite{xiao061},
there are 8 type diagrams contributing to the  $B_s \to \eta \eta^{'} $ decays, as illustrated in
Fig.1. We first calculate the usual factorizable diagrams (a) and (b). Operators $O_{1,2,3,4,9,10}$
are $(V-A)(V-A)$ currents, the sum of their amplitudes is given as
\beq
F_{e\eta}&=& 8\pi C_F m_{B_s}^4\int_0^1 d x_{1} dx_{3}\,
\int_{0}^{\infty} b_1 db_1 b_3 db_3\, \phi_{B_s}(x_1,b_1) \non & &
\times \left\{ \left[(1+x_3) \phi_\eta^A(x_3, b_3) +(1-2x_3) \re
(\phi_\eta^P(x_3,b_3) +\phi_\eta^T(x_3,b_3))\right] \right. \non
&& \left.\quad  \cdot \alpha_s(t_e^1)\,
h_e(x_1,x_3,b_1,b_3)\exp[-S_{ab}(t_e^1)] \right.\non && \left.
+2\re \phi_\eta^P (x_3, b_3)
\alpha_s(t_e^2)h_e(x_3,x_1,b_3,b_1)\exp[-S_{ab}(t_e^2)] \right\}.
\label{eq:ab}
\eeq
where $\re=m_0^\eta/m_B$; $C_F=4/3$ is a color factor. The explicit expressions of
the function $h_e$, the scales $t_e^i$ and the Sudakov factors $S_{ab}$ can be found Ref.~\cite{xiao061}.
The form factors of $B_s$ to $\eta$ decay, $F_{0,1}^{B_s \to \eta_{s\bar{s}}}(0)$, can thus be
extracted from the expression in Eq.~(\ref{eq:ab}).

\begin{figure}[t,b]
\vspace{-2 cm} \centerline{\epsfxsize=21 cm \epsffile{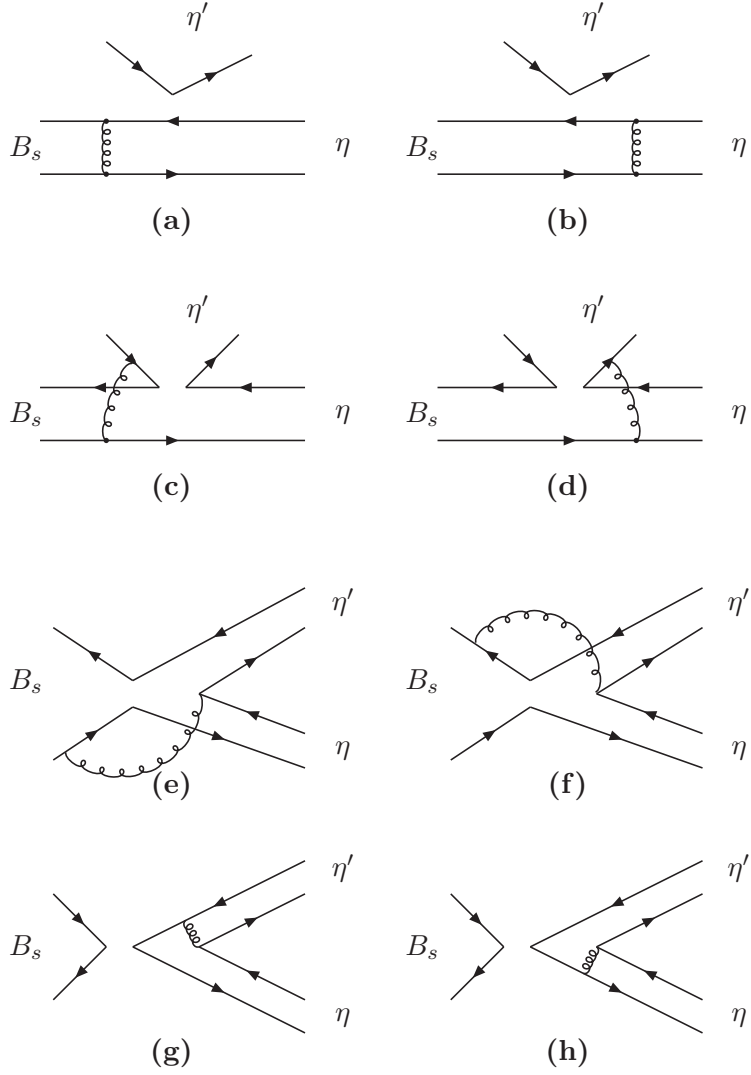}}
\vspace{-14cm} \caption{ Typical Feynman diagrams contributing to
the $B_s \to \eta\eta^{\prime}$  decays, where diagram (a) and (b)
contribute to the $B_s \to \eta$ form
factor $F_{0,1}^{B_s \to \eta}$.}
\label{fig:fig1}
\end{figure}

The operators $O_{5,6,7,8}$ have a structure of $(V-A)(V+A)$. Some
of these operators can contribute to the decay amplitude in a
factorizable way, but others may contribute after making a Fierz
transformation  in order to get right flavor and color structure
for factorization to work. Such kinds of contributions can be
written as \beq F_{e\eta}^{P1}&=&-F_{e\eta} \; .\\
F_{e\eta}^{P2}&=& 16\pi C_F m_{B_s}^4 \left
(\frac{(\fetap^s-\fetap^u)m_{\etapr}^2}{2 m_s m_{B_s}} \right ) \;
\int_{0}^{1}d x_{1}d x_{3}\,\int_{0}^{\infty} b_1d b_1 b_3d b_3\,
\pb(x_1,b_1) \non & & \times
 \left\{ \left[ \phi_\eta^A(x_3, b_3)+ \re((2+x_3) \pep (x_3, b_3)-x_3\pet(x_3, b_3))\right]
\right.  \non & &\left. \cdot \alpha_s (t_e^1)  h_e
(x_1,x_3,b_1,b_3)\exp[-S_{ab}(t_e^1)]\right.  \non & &\left. \
   +\left[x_1\pe(x_3, b_3)-2(x_1-1)\re\pep (x_3, b_3)\right]
\right.  \non & &\left. \cdot \alpha_s (t_e^2)
 h_e(x_3,x_1,b_3,b_1)\exp[-S_{ab}(t_e^2)] \right\} \; .
\eeq

For the non-factorizable diagrams 1(c) and 1(d), the corresponding decay amplitudes
can be written as
\beq
M_{e\eta}&=& \frac{16 \sqrt{6}}{3}\pi C_F m_{B_s}^4
\int_{0}^{1}d x_{1}d x_{2}\,d x_{3}\,\int_{0}^{\infty} b_1d b_1
b_2d b_2\, \pb(x_1,b_1) \pepr(x_2,b_2) \non
 & &\times
\left \{\left[2x_3\re \pet(x_3,b_1)-x_3 \phi_\eta(x_3,b_1)\right ]\right.\non
 & & \cdot \alpha_s(t_f) h_f(x_1,x_2,x_3,b_1,b_2)\exp[-S_{cd}(t_f)] \} \; , \\
M_{e\eta}^{P1}&=&0, \\
M_{e\eta}^{P2}&=& -M_{e\eta}\; .
\eeq

For the non-factorizable annihilation diagrams 1(e) and 1(f), we find
\beq
M_{a\eta}&=& \frac{16\sqrt{6}}{3}\pi C_F m_{B_s}^4\int_{0}^{1}d
x_{1}d x_{2}\,d x_{3}\,\int_{0}^{\infty} b_1d b_1 b_2d b_2\,
\phi_{B_s}(x_1,b_1)\non && \times \left \{ -\left \{x_2 \pe(x_3,
b_2) \pepr(x_2,b_2) \right.\right.\non
 & & \left.\left.
 +\re \rep \left [ (x_2+x_3+2) \peprp(x_2,b_2)
 +(x_2-x_3)\peprt(x_2,b_2) \right ] \pep(x_3,b_2)
\right.\right.\non
 & & \left.\left.
  +\re \rep \left [ (x_2-x_3)\peprp(x_3,b_2) +(x_2+x_3-2)\peprt(x_2,b_2)\right
  ]\pet(x_3,b_2)
\right\} \right. \non && \left. \quad \cdot
\alpha_s(t_f^3)h_f^3(x_1,x_2,x_3,b_1,b_2)\exp[-S_{ef}(t_f^3)]
\right.  \non && \left.
  + \left \{ x_3\pe(x_3,b_2) \pepr(x_2,b_2) \right.\right. \non
 && \left. \left.
 +\re \rep
 \left [ (x_2+x_3)\peprp(x_2,b_2)+(x_3-x_2)\peprt(x_2,b_2) \right ]\pep(x_3,b_2)
\right. \right. \non && \left.\left.
 +\re \rep\left [(x_3-x_2)\peprp(x_2,b_2)+(x_2+x_3)\peprt(x_2,b_2) \right ]\pet(x_3,b_2)
  \right \}
\right. \non && \left.
 \quad \cdot  \alpha_s(t_f^4)h_f^4(x_1,x_2,x_3,b_1,b_2)\exp[-S_{ef}(t_f^4) ]
 \right \}\; ,
 \eeq
 \beq
M_{a\eta}^{P1}&=& \frac{16\sqrt{6}}{3}\pi C_F
m_{B_s}^4\int_{0}^{1}d x_{1}d x_{2}\,d x_{3}\,\int_{0}^{\infty}
b_1d b_1 b_2d b_2\, \phi_{B_s}(x_1,b_1)
 \non
& &\times \left \{ \left[(x_3-2)
\re\pepr(x_2,b_2)(\pep(x_3,b_2)+\pet(x_3,b_2))-(x_2-2)\rep\pe(x_3,b_2)
\right.\right. \non && \left.\left.
(\peprp(x_2,b_2)+\peprt(x_2,b_2)) \right] \cdot
\alpha_s(t_f^3)h_f^3(x_1,x_2,x_3,b_1,b_2)\exp[-S_{ef}(t_f^3)]
\right. \non && \left.
 -\left[x_3\re\pepr(x_2,b_2)(\pep(x_3, b_2)+\pet(x_3,
b_2))\right. \right.\non && \left. \left.
 -x_2\rep\pe(x_3,b_2)(\peprp(x_2,b_2)+\peprt(x_2,b_2))\right]
\right. \non && \left.  \cdot
\alpha_s(t_f^4)h_f^4(x_1,x_2,x_3,b_1,b_2)\exp[-S_{ef}(t_f^4)]
\right \}
 \; ,
\eeq
\beq
 M_{a\eta}^{P2}&=& \frac{16\sqrt{6}}{3}\pi C_F m_{B_s}^4\int_{0}^{1}d x_{1}d
x_{2}\,d x_{3}\,\int_{0}^{\infty} b_1d b_1 b_2d b_2\,
\phi_{B_s}(x_1,b_1)\non && \times \left \{ \left \{x_3 \pe(x_3,
b_2) \pepr(x_2,b_2) \right.\right.\non
 & & \left.\left.
 +\re \rep \left [ (x_2+x_3+2) \peprp(x_2,b_2)
 +(x_3-x_2)\peprt(x_2,b_2) \right ] \pep(x_3,b_2)
\right.\right.\non
 & & \left.\left.
  +\re \rep \left [ (x_3-x_2)\peprp(x_3,b_2) +(x_2+x_3-2)\peprt(x_2,b_2)\right
  ]\pet(x_3,b_2)
\right\} \right. \non && \left. \quad \cdot
\alpha_s(t_f^3)h_f^3(x_1,x_2,x_3,b_1,b_2)\exp[-S_{ef}(t_f^3)]
\right.  \non && \left.
 -\left \{ x_2\pe(x_3,b_2) \pepr(x_2,b_2)
\right.\right. \non
 && \left. \left.
 +\re \rep
 \left [ (x_2+x_3)\peprp(x_2,b_2)+(x_2-x_3)\peprt(x_2,b_2) \right ]\pep(x_3,b_2)
\right. \right.\non && \left.\left.
 +\re \rep
 \left [(x_2-x_3)\peprp(x_2,b_2)+(x_2+x_3)\peprt(x_2,b_2) \right ]\pet(x_3,b_2)
 \right \}
\right. \non && \left.
 \quad \cdot  \alpha_s(t_f^4)h_f^4(x_1,x_2,x_3,b_1,b_2)\exp[-S_{ef}(t_f^4) ]
 \right \}\; . \label{eq:mapip2}
\eeq

For the factorizable annihilation diagrams 1(g) and 1(h), we have
\beq
F_{a\eta}&=& F_{a\eta}^{P1}= -8\pi C_F m_{B_s}^4\int_{0}^{1}dx_{2}\,d x_{3}\,
\int_{0}^{\infty} b_2d b_2b_3d b_3 \, \left\{ \left[x_3
\pe(x_3,b_3) \pepr(x_2,b_2)\right.\right.\non &&\left.\left.+2 \re
\rep((x_3+1)\pep(x_3,b_3)+(x_3-1) \pet(x_3,b_3))
\peprp(x_2,b_2)\right] \right. \non && \left. \quad \cdot
\alpha_s(t_e^3) h_a(x_2,x_3,b_2,b_3)\exp[-S_{gh}(t_e^3)] \right.
\non && \left. -\left[ x_2 \pe(x_3,b_3) \pepr(x_2,b_2)
\right.\right.\non && \left. \left. \quad +2 \re \rep
((x_2+1)\peprp(x_2,b_2)+(x_2-1) \peprt(x_2,b_2)
)\pep(x_3,b_3)\right] \right. \non &&\left. \quad \cdot
\alpha_s(t_e^4)
 h_a(x_3,x_2,b_3,b_2)\exp[-S_{gh}(t_e^4)]\right \}\; \\
F_{a\eta}^{P2}&=& -16 \pi C_F m_{B_s}^4 \int_{0}^{1}d
x_{2}\,d x_{3}\,\int_{0}^{\infty} b_2d b_2b_3d b_3 \,\non &&\times
\left\{ \left[x_3 \re (\pep(x_3, b_3)-\pet(x_3,
b_3))\pepr(x_2,b_2)+2\rep \pe(x_3,b_3) \peprp(x_2,b_2)
\right]\right.
 \non
&&\left.\quad \cdot \alpha_s(t_e^3)
h_a(x_2,x_3,b_2,b_3)\exp[-S_{gh}(t_e^3)]\right.
 \non
 &&\left.+\left[x_2\rep(\peprp(x_2,b_2)-\peprt(x_2,b_2))\pe(x_3,b_3)+2\re
\pepr(x_2,b_2)\pep(x_3,b_3)\right] \right.\non &&\left.\quad \cdot
 \alpha_s(t_e^4)
 h_a(x_3,x_2,b_3,b_2)\exp[-S_{gh}(t_e^4)]\right\}\;.
 \eeq

For the $B_s \to \eta\eta^{\prime}$ decay, besides the Feynman
diagrams as shown in Fig.~1 where the upper emitted meson is the
$\etapr$, the Feynman diagrams obtained by exchanging the position
of $\eta$ and $\etapr$ also contribute to this decay mode. The
corresponding expressions of amplitudes for new diagrams will be
similar with those as given in Eqs.(\ref{eq:ab}-\ref{eq:mapip2}),
since the $\eta$ and $\etapr$ are all light pseudoscalar mesons
and have the similar wave functions. The expressions of amplitudes
for new diagrams can be obtained by the replacements \beq \pe
\longleftrightarrow  \phi_{\etapr},   \quad
\pep\longleftrightarrow  \phi^P_{\etapr}, \quad \pet
\longleftrightarrow \phi^t_{\etapr}, \quad \re \longleftrightarrow
r_{\etapr}. \eeq

 For example, we find that:
 \beq
  F_{e\etapr}&=&F_{e\eta},\quad F_{a\etapr}=-F_{a\eta},
  \quad F_{a\etapr}^{P1}=-F_{a\eta}^{P1},\quad
 F_{a\etapr}^{P2}=F_{a\eta}^{P2}.
 \eeq

%%%%%%%%%%%%%%%%%%%%%%%%%%%%%%%%%%%%%%%%%%%%%%%%%%%%%%%%%%%%%%%%%

Before we write down the complete decay amplitude for the studied
decay modes, we firstly give a brief discussion about the
$\eta-\eta^\prime$ mixing and the gluonic component of the
$\eta^\prime$ meson. There exist two popular mixing basis for
$\eta-\eta^\prime$ system, the octet-singlet and the quark flavor
basis, in literature. Here we  use the $SU(3)_F$ octet-singlet
basis with the two mixing angle ($\theta_1, \theta_8$)
scheme \cite{fk98} to describe the mixing of
$\eta$ and $\eta^\prime$ mesons.
In the numerical calculations, we will use the following
mixing parameters \cite{fk98}
\beq
\theta_8 &=&-21.2^\circ, \quad \theta_1=-9.2^\circ, \quad f_1= 1.17 f_\pi,
\quad f_8 = 1.26 f_\pi .   \label{eq:t1-t8}
\eeq

In this paper, we firstly take $\eta$ and $\eta^\prime$ as a linear combination
of light quark pairs $u\bar{u}, d\bar{d}$ and $s\bar{s}$, and then
estimate the possible gluonic contributions to $B_s \to \etap \etap$ decays
by using the formulae as presented in Ref.~\cite{li0609}.
We found that the possible gluonic contributions are indeed small.

%%%%%%%%%%%%%%%%%%%%%%%%%%%%%%%%%%%%%%%%%%%%%%%%%%%%%%%%%%%%%%%

\subsection{Complete decay amplitudes}

For $B_s^0 \to \eta \etapr$ decay, by combining the contributions
from different diagrams, the total decay amplitude can be written as
\beq
{\cal M}(\eta\etapr) &=& \left
 (F_{e\eta} F_2 f_{\eta^\prime}^d +F_{e\etapr}F'_2 f_\eta^d\right )
 \cdot\left[ \xi_u \left( C_1 +
\frac{1}{3}C_2\right)\right.
 \non
& &\left.-\xi_t
\left(C_{3}+\frac{1}{3}C_{4}-C_{5}-\frac{1}{3}C_{6}+
\frac{1}{2}C_7+\frac{1}{6}C_8-\frac{1}{2}C_9-\frac{1}{6}C_{10}\right)\right
]
 \non
& &- \left( F_{e\eta} F_2 f_{\eta^\prime}^s
+F_{e\etapr} F'_2 f_\eta^s\right )\non & &
\cdot\,\xi_t\left(\frac{7}{3}C_3+\frac{5}{3}C_4-2C_{5}-\frac{2}{3}C_{6}
-\frac{1}{2}C_7-\frac{1}{6}C_8+\frac{1}{3}C_9
 -\frac{1}{3} C_{10}\right) \non && -
 \left(F_{e\eta}^{P_2}F_2
+F_{e\etapr}^{P_2}F'_2 \right)\xi_t \left
(\frac{1}{3}C_5+C_6 -\frac{1}{6}C_7-\frac{1}{2}C_{8}\right) \non
&& +\left( M_{e\eta}+
M_{e\etapr}\right)F_2 F'_2 \left
[ \xi_uC_2-\xi_t \left(C_3+2C_4-\frac{1}{2}C_9
+\frac{1}{2}C_{10}\right)\right ]
  \non &&  -\left[ M_{e\eta}F_1 F_2' +
M_{e\etapr} F'_1 F_2 \right]
\xi_t \left ( C_4 -\frac{1}{2}C_{10}\right )
  \non && -\left( M_{e\eta}^{P_2}+
  M_{e\etapr}^{P_2}\right)\,F_2 F'_2 \xi_t\,\left(2C_6+\frac{1}{2}C_8\right)\non && -
\left(M_{e\eta}^{P_2}F_1 F_2' + M_{e\etapr}^{P_2} F'_1 F_2 \right)\xi_t\,
\left(C_6-\frac{1}{2}C_8\right)\non && +
\left(M_{a\eta}+M_{a\etapr}\right)
F_1 F'_1 \left[ \xi_uC_{2}-\xi_t \left(C_3+2C_4-\frac{1}{2}C_9
+\frac{1}{2}C_{10}\right)\right ] \non &&
 -\left(M_{a\eta}+M_{a\etapr}\right) F_2 F'_2 \xi_t \left ( C_4 -\frac{1}{2}C_{10}\right ) \non &&
-\left(M_{a\eta}^{P_1}\,+M_{a\etapr}^{P_1}\,\right) F_2 F'_2
\xi_t\,\left(C_{5}-\frac{1}{2}C_{7}\right)
 \non && -
 \left(M_{a\eta}^{P_2}\,+
 M_{a\etapr}^{P_2}\,\right) F_1 F'_1 \xi_t\,
\left(2C_6+\frac{1}{2}C_8\right) \non && -
 \left(M_{a\eta}^{P_2}\,+
 M_{a\etapr}^{P_2}\,\right) F_2 F'_2 \xi_t\,  \left(C_6-\frac{1}{2}C_8\right)
\non &&-f_{B_s}\cdot\left(F_{a\eta}^{P2}+
F_{a\etapr}^{P2}\right) F_2 F'_2 \xi_t \left (\frac{1}{3}C_5+C_6 -\frac{1}{6}C_7-\frac{1}{2}C_{8}\right)
, \label{eq:m1}
\eeq
where $\xi_u = V_{ub}^*V_{us}$ and $\xi_t = V_{tb}^*V_{ts}$, and
the relevant mixing parameters and decay constants are
\beq
F_1 &=& \sqrt{\frac{1}{6}}\cos\theta_8 - \sqrt{\frac{1}{3}} \sin\theta_1,
\quad
F_2=-\sqrt{\frac{2}{3}}\sin\theta_8 + \sqrt{\frac{1}{3}} \cos\theta_1,\label{eq:f1f2} \\
F'_1 &=& \sqrt{\frac{1}{6}} \sin{\theta_8} + \sqrt{\frac{1}{3}}
\cos{\theta_1}, \quad
F'_2= -\sqrt{\frac{2}{3}} \sin{\theta_8}  + \sqrt{\frac{1}{3}} \cos{\theta_1}, \\
f_{\eta}^d &=&  \frac{f_8}{\sqrt{6}} \cos\theta_8 -
\frac{f_1}{\sqrt{3}}\sin\theta_1, \quad f_{\eta}^s = -\frac{2
f_8}{\sqrt{3}} \cos\theta_8  - \frac{f_1}{\sqrt{3}} \sin\theta_1, \\
f_{\eta^\prime}^d &=& \frac{f_8}{\sqrt{6}} \sin\theta_8 +
\frac{f_1}{\sqrt{3}}\cos\theta_1, \quad f_{\eta^\prime}^s =
-\frac{2 f_8}{\sqrt{3}} \sin\theta_8 + \frac{f_1}{\sqrt{3}}
\cos\theta_1 .
\eeq

Similarly, the decay amplitudes for $B_s^0 \to \eta \eta$ and
 $B_s^0 \to \eta^{\prime} \eta^{\prime}$
decay can be obtained easily from Eq.(\ref{eq:m1})  by the
following replacements
\beq
f_\eta^{d},\; f_\eta^s
\longleftrightarrow  f_{\eta^\prime}^d, \; f_{\eta^\prime}^s;
\quad F_1(\theta_1,\theta_8) \longleftrightarrow
F'_1(\theta_1,\theta_8); \quad F_2(\theta_1,\theta_8)
\longleftrightarrow F'_2(\theta_1,\theta_8).
\eeq
Note that the contributions from the possible gluonic component of $\eta'$ meson
have not been included here.

\section{Numerical results and Discussions}\label{sec:n-d}

In this section, we will calculate the CP-averaged branching ratios and CP
violating asymmetries for those considered decay modes. The input
parameters and the wave functions to be used are given in Appendix
\ref{sec:app2}. In numerical calculations, central values of input
parameters will be used implicitly unless otherwise stated.

Using the decay amplitudes obtained in last section, it is
straightforward to calculate the branching ratios.
By employing the two mixing angle scheme of $\eta-\eta^\prime$
system and using the mixing parameters as given in
Eq.~(\ref{eq:t1-t8}), one finds the CP-averaged  branching ratios
for the considered three decays as follows
\beq
Br(\ B_s^0 \to\eta \eta) &=& \left [14.2^{+6.6}_{-4.2}(\omega_b)
 ^{+16.7}_{-6.2}(m_0^{\eta_{s\bar{s}}}) \right ] \times 10^{-6}, \label{eq:bree1}\\
Br(\ B_s^0 \to \eta\eta^{\prime}) &=& \left [12.4^{+5.7}_{-3.6}( \omega_b)
^{+17.3}_{-6.0}(m_0^{\eta_{s\bar{s}}}) \right ] \times 10^{-6}, \label{eq:brep1} \\
Br(\ B_s^0 \to \eta^{\prime}\eta^{\prime}) &=& \left [9.2 ^{+3.0} _{-2.0}
(\omega_b)^{+15.0}_{-4.5}(m_0^{\eta_{s\bar{s}}}) \right ]\times 10^{-6},
 \label{eq:brpp1}
\eeq
where the main errors are induced by the uncertainties of $\omega_b=0.50 \pm 0.05$ GeV,
and $m_0^{\eta_{s\bar{s}}} = [1.49-2.38]$ GeV  (corresponding to $m_s=130\pm 30$ MeV),
respectively. The above pQCD predictions agree well with those obtained in the QCD facterization approach
\cite{sunbs03}.

As for the gluonic contributions, we follow the same procedure as being used in
Ref.~\cite{li0609} to include the possible gluonic contributions
to the $B_s \to \etap$ transition form factors $F_{0,1}^{B_s \to
\etap}$ and found that the gluonic contributions to the branching
ratios are less than $3\%$ for $B \to \eta \eta$ decay, $\sim
7\%$ for $B \to \eta \eta^\prime$ decay, and around $18\%$ for $B
\to \eta^\prime \eta^\prime$ decay.
The central values of the pQCD
predictions for $B_s \to \etap \etap$ decays after the inclusion
of possible gluonic contributions are the following
\beq
Br( B_s^0 \to \eta \eta) &=& \left [13.7^{+6.4}_{-4.0}(\omega_b)
 ^{+16.5}_{-6.1}(m_0^{\eta_{s\bar{s}}})\right ] \times 10^{-6}, \label{eq:breeg1}\non
Br(\ B_s^0 \to \eta\eta^{\prime}) &=& \left [11.6^{+5.3}
_{-3.4}( \omega_b)^{+16.8}_{-5.7}(m_0^{\eta_{s\bar{s}}}) \right ] \times 10^{-6}, \label{eq:brepg1} \non
Br(\ B_s^0 \to \eta^{\prime}\eta^{\prime}) &=& \left [10.8^{+3.7} _{-2.4}
(\omega_b)^{+16.2}_{-5.2}(m_0^{\eta_{s\bar{s}}}) \right ]\times 10^{-6}.
 \label{eq:brppg1}
\eeq

Now we turn to the evaluations of the CP-violating asymmetries of $B_s \to \etap \etap$
decays in pQCD approach.
For $B_s^0$ meson decays, a non-zero ratio $(\Delta \Gamma/\Gamma)_{B_s}$ is expected in the SM
\cite{bz99,f2006}. For $B_s \to \etap \etap$ decays, three quantities to describe the CP violation
can be defined as follows \cite{f2006}:
\beq
\acp^{dir}=\frac{ \left | \lambda_{CP}\right |^2 -1 } {1+|\lambda_{CP}|^2}, \quad
\acp^{mix}=\frac{ 2Im ( \lambda_{CP})}{1+|\lambda_{CP}|^2}, \quad
{\cal A}_{\Delta \Gamma_s}= \frac{ 2 Re ( \lambda_{CP})}{1+|\lambda_{CP}|^2},
\label{eq:acp-dm}
\eeq
with
\beq
\lambda_{CP} =\eta_f \; \frac{ V_{tb}^*V_{ts} \langle f |H_{eff}| \bar{B_s^0}\rangle} {
V_{tb}V_{ts}^* \langle f |H_{eff}| B_s^0\rangle} \approx
\frac{ \langle f |H_{eff}| \bar{B_s^0}\rangle} {\langle f |H_{eff}| B_s^0\rangle} ,
 \label{eq:lambda2}
\eeq
in a very good approximation. Here $\acp^{dir}$ and $\acp^{mix}$ means the direct and
mixing-induced CP violation respectively,
while the third term ${\cal A}_{\Delta \Gamma_s}$
is related to the presence of a non-negligible $\Delta \Gamma_s$.
By using the mixing parameters in Eq.~(\ref{eq:t1-t8}) and the input parameters as given in Appendix A,
one found the pQCD predictions for $\acp^{dir}$, $\acp^{mix}$ and $H_f$
\beq
\acp^{dir}(B_s^0 \to\eta \eta) &=&  \left [ -0.2 \pm 0.1 (\gamma) \pm 0.1 (\omega_b) ^{+0.4}_{-0.2}
(m_0^{\eta_{s\bar{s}}})\right ] \times 10^{-2}, \non
\acp^{dir}(B_s^0 \to \eta \eta^\prime) &=& \left [ + 0.6^{+0.1}_{-0.2}
(\gamma) \pm 0.1 (\omega_b) \pm 0.3 (m_0^{\eta_{s\bar{s}}}) \right] \times 10^{-2}, \non
\acp^{dir}(B_s^0 \to \eta^\prime \eta^\prime) &=& \left [ -0.8 ^{+0.2}_{-0.1} (\gamma)
\pm 0.1 (\omega_b) \pm 0.7 (m_0^{\eta_{s\bar{s}}}) \right ] \times 10^{-2},\\
%%%%
\acp^{mix}(B_s^0 \to\eta \eta) &=& \left [ -0.3 \pm 0.1 (\gamma) \pm 0.2(\omega_b)
\pm 0.5 (m_0^{\eta_{s\bar{s}}}) \right ]\times 10^{-2},\non
\acp^{mix}(B_s^0\to \eta \eta^\prime) &=& \left [ -0.8\pm 0.2 (\gamma) \pm 0.1(\omega_b)
\pm 0.2 (m_0^{\eta_{s\bar{s}}}) \right ]\times 10^{-2},  \non
\acp^{mix}(B_s^0 \to \eta^\prime \eta^\prime) &=& \left [ +1.8
^{+0.3}_{-0.5} (\gamma) \pm 0.0 (\omega_b) ^{+0.5}_{-0.3} (m_0^{\eta_{s\bar{s}}})
\right ]\times 10^{-2}, \\
{\cal A}_{\Delta \Gamma_s}(\eta \eta)&\approx & {\cal A}_{\Delta \Gamma_s}(\eta \eta^\prime) \approx
{\cal A}_{\Delta \Gamma_s}(\eta^\prime \eta^\prime)\approx 1 ,
\eeq
where the dominant errors come from the variations of CKM angle
$\gamma=60^\circ \pm 20^\circ$, $\omega_b=0.50\pm 0.05$ GeV and
$m_0^{\eta_{s\bar{s}}} = [1.49-2.38]$ GeV ( corresponding to $m_s=130\pm 30$ MeV), respectively.
It is easy to see that both the direct and mixing-induced CP violations of the considered $B_s$
decays are very small in magnitude, and thus almost impossible to measure them even in
the LHC experiments. The above pQCD predictions are also
consistent with the QCDF predictions \cite{bn03b,sunbs03}.

In short, we calculated the branching ratios and
CP-violating asymmetries of  $B_s^0 \to \eta \eta$, $ \eta
\eta^\prime$ and $ \eta^{\prime}\eta^{\prime}$ decays at the
leading order by using the pQCD factorization approach.
Besides the usual factorizable diagrams, the non-factorizable and
annihilation diagrams are also calculated analytically in the pQCD
approach. From our calculations and phenomenological analysis, we found the
following results:
\begin{itemize}
\item
Using the two mixing angle scheme, the pQCD predictions for
the CP-averaged branching ratios are
\beq
Br(B_s^0 \to \eta \eta)
&=& \left (14.2 ^{+18.0}_{-7.5} \right ) \times 10^{-6},\non
Br(B_s^0 \to \eta \eta^\prime)&=& \left ( 12.4 ^{+18.2}_{-7.0}
\right ) \times 10^{-6},\non
Br(B_s^0 \to \eta^{\prime} \eta^{\prime}) &=& \left (
9.2^{+15.3}_{-4.9} \right ) \times 10^{-6},
\eeq
where the various errors as specified previously have
been added in quadrature. The pQCD predictions for the three decay
channels agree well with those obtained by employing the QCDF
approach.

\item
The gluonic contributions are small in size: less than $7\%$
for $B_s \to \eta \eta$ and  $ \eta \eta^\prime$ decays,
and around $18\%$ for $B_s \to \eta' \eta'$ decay.

\item
The direct and mixing-induced CP violations of the considered three
decay modes are very small: less than $3\%$ in magnitude.

\end{itemize}

\noindent
Note added: After completion of this paper, the paper in Ref.\cite{ali07} appeared, and where
a systematic study for the $B_s \to M_1 M_2$ decays in the pQCD factorization approach has been done.
Since different mixing-scheme of $\eta-\eta^\prime$ system have been used, the explicit
expressions of the decay amplitudes of the relevant decays are different in these two papers,
but the numerical predictions for branching ratios and CP violations agree well with
each other. The possible gluonic contributions are estimated here.

\begin{acknowledgments}

X.~Liu would like to acknowledge the financial support of The Scientific Research
Start-up Fund of Zhejiang Ocean University under Grant No.21065010706.
This work was partially supported by the National Natural Science Foundation of China under Grant
No.10575052, and by the Specialized Research Fund for
the Doctoral Program of Higher Education (SRFDP) under Grant No.~20050319008.

\end{acknowledgments}

%%%%%%%%%%%%%%%%%%%%%%%%%%%%%%%%%%%%%%%%%%%%%%%%%%%%%%%%%%%%%%%%%%%%%%%%%%%%%%%%%%
%                                        Appendix
%%%%%%%%%%%%%%%%%%%%%%%%%%%%%%%%%%%%%%%%%%%%%%%%%%%%%%%%%%%%%%%%%%%%%%%%%%%%%%%%5

\begin{appendix}

\section{Input parameters and wave functions} \label{sec:app2}

In this Appendix we show the input parameters and the light meson
wave functions to be used in the numerical calculations.

The masses, decay constants, QCD scale  and $B_s^0$ meson lifetime
are \beq
 \Lambda_{\overline{\mathrm{MS}}}^{(f=4)} &=& 250 {\rm MeV}, \quad
 f_\pi = 130 {\rm MeV}, \quad  f_{B_s} = 230 {\rm MeV}, \non
 m_0^{\eta_{d\bar{d}}}&=& 1.4 {\rm GeV},\quad m_s=130 {\rm
 MeV},  \quad
  f_K = 160  {\rm MeV}, \non
 M_{B_s} &=& 5.37 {\rm GeV}, \quad M_W = 80.41{\rm
 GeV},\tau_{B_s^{0}}=1.46\times10^{-12}{\rm s}
 \label{para}
\eeq

For the CKM matrix elements, here we adopt the Wolfenstein
parametrization for the CKM matrix, and take $\lambda=0.2272,
A=0.818, \rho=0.221$ and $\eta=0.340$ \cite{pdg06}.

For the $B$ meson wave function, we adopt the model \beq
\phi_{B_s}(x,b) &=& N_{B_s} x^2(1-x)^2 \mathrm{exp} \left
 [ -\frac{M_{B_s}^2\ x^2}{2 \omega_{b}^2} -\frac{1}{2} (\omega_{b} b)^2\right],
 \label{phib}
\eeq where $\omega_{b}$ is a free parameter and we take
$\omega_{b}=0.50\pm 0.05$ GeV in numerical calculations, and
$N_{B_s}=63.67$ is the normalization factor for $\omega_{b}=0.50$.

For the distribution amplitudes $\phi_{\eta_{d\bar{d}}}^A$,
$\phi_{\eta_{d\bar{d}}}^P$ and $\phi_{\eta_{d\bar{d}}}^T$ , we
utilize the result from the light-cone sum rule~\cite{ball}
including twist-3 contribution. For the corresponding Gegenbauer moments and relevant input parameters,
we here use $a^{\eta_{d\bar{d}}}_2= 0.115, a^{\eta_{d\bar{d}}}_4=-0.015$,
$\rho_{\eta_{d\bar{d}}}= m_{\pi}/{m_0^{\eta_{d\bar{d}}}}$,
$\eta_3=0.015$ and $\omega_3=-3.0$.
We also assume that the wave function of $u\bar{u}$ is the same as the
wave function of $d\bar{d}$ \cite{ekou}. For the wave function of
the $s\bar{s}$ components, we also use the same form as $d\bar{d}$
but with $m^{s\bar{s}}_0$ and $f_y$ instead of $m^{d\bar{d}}_0$
and $f_x$, respectively:
 \beq
 f_x=f_{\pi}, \ \ \ f_y=\sqrt{2f_K^2-f_{\pi}^2}.\ \ \
\label{eq:fxfy}
\eeq
These values are translated to the values in the two mixing angle
method:
\beq
f_1&=& 152.1  {\rm MeV}, \quad f_8 =163.8  {\rm MeV}, \non
\theta_1 &= & -9.2^\circ, \quad \theta_8= -21.2^{\circ}. \quad
\eeq
The parameters $m_0^i$ $(i=\eta_{d\bar{d}(u\bar{u})},
\eta_{s\bar{s}})$ are defined as:
\beq
m_0^{\eta_{d\bar{d}(u\bar{u})}}\equiv m_0^\pi \equiv
\frac{m_{\pi}^2}{(m_u+m_d)}, \qquad m_0^{\eta_{s\bar{s}}}\equiv
\frac{2M_K^2-m_{\pi}^2}{(2m_s)}.
\eeq

\end{appendix}

%%%%%%%%%%%%%%%%%%%%%%%%%%%%%%%%%%%%%%%%%%%%%%%%%%%%%%%%%%%%%%%%%%%%%%%%%%%%%%%%%%%%%%%%%%%%%%5
%                                 reference
%%%%%%%%%%%%%%%%%%%%%%%%%%%%%%%%%%%%%%%%%%%%%%%%%%%%%%%%%%%%%%%%%%%%%%%%%%%%%%%%%%%%%%%%%%%%%%%%%

\end{document}